\documentclass[a4paper,useAMS,usenatbib]{mn2e}
\usepackage[final]{graphics}
\usepackage{epsfig}
\usepackage{amssymb}

\newcommand{\he}[1] {He\,{\sc #1}}
\newcommand{\hel}[2] {He\,{\sc #1}~$\lambda$#2}
\newcommand{\kms}{\mbox{$\mathrm{km~s^{-1}}$}}

\newcommand{\ion}[2] {#1\,{\sc #2}}

\title[The 98.10-min orbital period of nova V458 Vul]{The orbital period of V458 Vulpeculae, a post double common-envelope nova}
\author[P. Rodr\'{i}guez-Gil et al.]{P. Rodr\'{i}guez-Gil$^{1,2,3}$\thanks{E-mail: prguez@ing.iac.es (PRG)}, M. Santander-Garc\'{i}a$^{1,2,3}$, C. Knigge$^{4}$, R. L. M. Corradi$^{2,3}$,
\newauthor
B. T. G\"ansicke$^{5}$, M. J. Barlow$^{6}$, J. J. Drake$^{7}$, J. Drew$^{8}$, B. Miszalski$^{8}$,
\newauthor
R. Napiwotzki$^{8}$, D. Steeghs$^{5}$, R. Wesson$^{6}$, A. A. Zijlstra$^{9}$, D. Jones$^{9}$, T. Liimets$^{10,1}$, 
\newauthor
S. Pyrzas$^{1,5}$ and M. M. Rubio-D\'\i ez$^{1,11}$\\   
$^{1}$Isaac Newton Group of Telescopes, Apartado de Correos 321, Santa Cruz de La Palma, E-38700, Spain\\
$^{2}$Instituto de Astrof\'\i sica de Canarias, V\'\i a L\'actea, s/n, La Laguna, E-38205, Santa Cruz de Tenerife, Spain\\
$^{3}$Departamento de Astrof\'\i sica, Universidad de La Laguna, La Laguna, E-38205, Santa Cruz de Tenerife, Spain\\
$^{4}$School of Physics and Astronomy, University of Southampton, Southampton SO17 1BJ, UK\\
$^{5}$Department of Physics, University of Warwick, Coventry CV4 7AL, UK\\
$^{6}$Department of Physics and Astronomy, University College London, Gower Street, London WC1E 6BT, UK\\
$^{7}$Harvard-Smithsonian Center for Astrophysics, 60 Garden Street, Cambridge, MA 02138, USA\\
$^{8}$Centre for Astrophysics Research, STRI, University of Hertfordshire, College Lane Campus, Hatfield AL10 9AB, UK\\
$^{9}$Jodrell Bank Center for Astrophysics, Alan Turing Building, The University of Manchester, Oxford Road, Manchester M13 9PL, UK\\
$^{10}$Tartu Observatoorium, T\~oravere 61602, Estonia\\
$^{11}$Centro de Astrobiolog\'\i a (CSIC/INTA), Ctra. de Torrej\'on a Ajalvir, km 4, E-28850 Torrej\'on de Ardoz, Madrid, Spain
}
\begin{document}
\date{Accepted 2010. Received 2010}
\pagerange{\pageref{firstpage}--\pageref{lastpage}} \pubyear{2010}
\maketitle
\label{firstpage}

\begin{abstract}

We present time-resolved optical spectroscopy of V458 Vulpeculae (Nova Vul 2007 No. 1) spread over a period of 15 months starting 301 days after its discovery. Our data reveal radial velocity variations in the \hel{ii}{5412} and \hel{ii}{4686} emission lines. A period analysis of the radial velocity curves resulted in a period of $98.09647\pm0.00025$~min ($0.06812255 \pm 0.00000017$~d) which we identify with the orbital period of the binary system. V458 Vul is therefore the planetary nebula central binary star with the shortest period known. We explore the possibility of the system being composed of a relatively massive white dwarf ($M_1 \gtrsim 1.0~\mathrm{M_\odot}$) accreting matter from a post-asymptotic giant branch star which produced the planetary nebula observed. In this scenario, the central binary system therefore underwent two common-envelope episodes. A combination of previous photoionisation modelling of the nebular spectra, post-asymptotic giant branch evolutionary tracks and the orbital period favour a mass of $M_2 \sim 0.6~\mathrm{M_\odot}$ for the donor star. Therefore, the total mass of the system may exceed the Chandrasekhar mass, which makes V458 Vul a Type Ia supernova progenitor candidate.    

\end{abstract}


\section{Introduction}

\begin{table*}
\label{t-obslog}
\setlength{\tabcolsep}{0.95ex}
\caption[]{Log of the observations.}
\vspace*{-2.5ex}

\begin{tabular}[t]{lcccccccc}
\hline\noalign{\smallskip}
Date & Telescope/  &  Grating & Slit width & Wavelength      & Exp. time & Time  coverage               &  Dispersion  & Resolution   \\    
         &  Instrument  &                 & (arcsec)  & range      & (s)             &  (h)  &  (\AA~pix$^{-1}$) & (\AA)\\    
\hline\noalign{\smallskip}
2008 Jun 04 & INT/IDS            & R300V   & 1.2         & $\lambda\lambda$3320--8400    & 300 & 5.7    &  1.9  & 5.0 \\
2008 Jun 25 &  INT/IDS           & R632V   & 1.5         & $\lambda\lambda$4405--7150    & 300 & 3.3    & 0.9   &  2.8\\
2008 Jul 04 & WHT/ISIS          & R1200B & 1.0         & $\lambda\lambda$4929--5644    & 300 & 7.4    & 0.2   & 0.7 \\
2008 Oct 13 &  INT/IDS            & R300V & 1.0         & $\lambda\lambda$3288--8870        & 300 & 3.6    & 1.9   &  4.4 \\
2008 Oct 14 &  INT/IDS            & R300V  & 1.0          & $\lambda\lambda$3294--8400       &  300 & 3.6     & 1.9   & 4.4 \\
2008 Nov 11 & WHT/ISIS        & R600B  & 1.0       & $\lambda\lambda$3584--5117           &  300 & 3.7    & 0.9   & 1.5  \\
2009 May 25 &  INT/IDS          & R632V   & 1.2        & $\lambda\lambda$4500--6830    &  600 &  3.2   &  0.9  &  2.4 \\
2009 Jul 21  & WHT/ISIS   &  R600B &  1.0  & $\lambda\lambda$4500--4930  &  30  & 4.4 &0.4 & 1.5\\
  & (QUCAM3)   &   &    &   &    &  & & \\
2009 Aug 31 &  INT/IDS          & R632V   & 1.2        & $\lambda\lambda$4500--6830    &  600, 800 &  4.0   &  0.9  &  2.4 \\
\noalign{\smallskip}\hline
\end{tabular}

\smallskip
\begin{minipage}{158mm}
Notes on instrumentation: 
INT/IDS: 2.5-m
Isaac Newton Telescope at Roque de los Muchachos Observatory (ORM), using the Intermediate 
Dispersion Spectrograph with a $2\mathrm{k}\times4\mathrm{k}$ pixel E2V CCD; 
WHT/ISIS: 4.2-m William Herschel Telescope (WHT) at ORM, 
using the Intermediate dispersion Spectrograph and Imaging System with its $2\mathrm{k}\times4\mathrm{k}$ pixel E2V 
CCD.
\end{minipage}
\end{table*}

V458 Vul (Nova Vul 2007 No. 1) was discovered at 9.5 magnitude on 2007 August 8 \citep{nakanoetal07-1}, shortly before peaking at $V=8.1$. It is classified as a fast nova on the basis of its rapid three-magnitude brightness fall from maximum within 21 days, indicative of a relatively massive ($\sim 1~\mathrm{M_\odot}$) white dwarf. In our first paper \citep[][hereafter W08]{wessonetal08-1} we reported the discovery of a wasp-waisted planetary nebula surrounding the $r^{\prime} = 18.34$ nova progenitor, and speculated about the possibility of the central binary star in V458 Vul being composed of a white dwarf and a post-asymptotic giant branch (post-AGB) star which formed the planetary nebula. However, the lack of an accurate orbital period prevented any further discussion. In addition, \cite{goranskijetal08-1} had suggested a tentative orbital period of 0.59 d from photometric light curves. In an attempt to measure a precise orbital period we started a time-resolved spectroscopy campaign searching for the orbital signature in the radial velocities of the emission lines. The results of this campaign are presented in this letter.




\section{Observations and data reduction}

The spectroscopic data were obtained with the Intermediate Dispersion Spectrograph (IDS) on the 2.5-m Isaac Newton Telescope (INT) and  the  Intermediate dispersion Spectrograph and Imaging System (ISIS) on the 4.2-m William Herschel Telescope (WHT), both on La Palma. The log of spectroscopic observations can be found in Table~\ref{t-obslog}.

The spectra were reduced using the standard \texttt{IRAF} long-slit packages. The one-dimensional spectra were then extracted using the optimal extraction algorithm of \cite{horne86-1}. Wavelength calibration was performed in \texttt{MOLLY}\footnote{http://www.warwick.ac.uk/go/trmarsh/software} by means of arc lamp spectra frequently taken to guarantee an accurate wavelength solution. The spectra were then flux calibrated and de-reddened using $E(B-V)=0.63$ (W08) using \texttt{MOLLY}. For the fast spectroscopy QUCAM3 data we averaged in blocks of 10 spectra in order to achieve a proper signal-to-noise ratio for radial velocity measurement.

\section{Early radial velocity variability}

\begin{figure}
\includegraphics[width=\columnwidth]{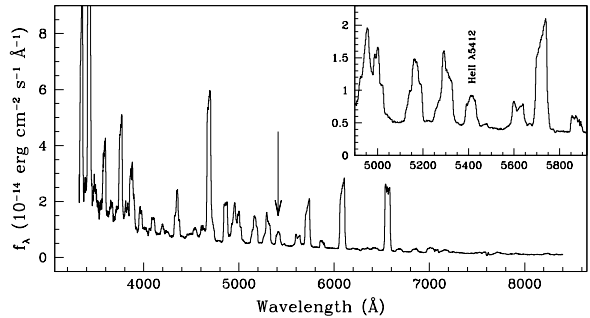}
\caption{Average of 15 spectra taken with INT/IDS on 2008 June 4 (301 days after the nova explosion). The arrow points to the \hel{ii}{5412} emission line, which showed significant radial velocity shifts while the line emission was still dominated by the nova shell.}
\label{fig_inispec}
\end{figure}

The average optical spectrum of V458 Vul taken on 2008 June 4 (day 301 after the nova explosion) is shown in Fig.~\ref{fig_inispec}. It is mainly dominated by emission lines of [\ion{Ne}{v}], [\ion{Fe}{vii}], \ion{He}{ii}, and the hydrogen Balmer series. The round-topped profile of the \hel{ii}{5412} line attracted our attention during a first visual inspection of the line shapes. After normalising the adjacent continuum, we cross-correlated the individual \hel{ii}{5412} profiles with a Gaussian template with a full-width at half-maximum (FWHM) of 400 \kms. We found the radial velocity to vary between $\sim -200$ and $\sim 200$ \kms. This radial velocity variation indicated that at least one of the components of the \hel{ii}{5412} emission forms in a binary system at the core of the planetary nebula. This finding prompted further time-resolved spectroscopy (see Table~\ref{t-obslog}) in an attempt to measure its orbital period.      

\section{The orbital period of V458 Vul}

\subsection{Period analysis of the radial velocity curves}

\begin{figure}
\includegraphics[height=\columnwidth,angle=-90]{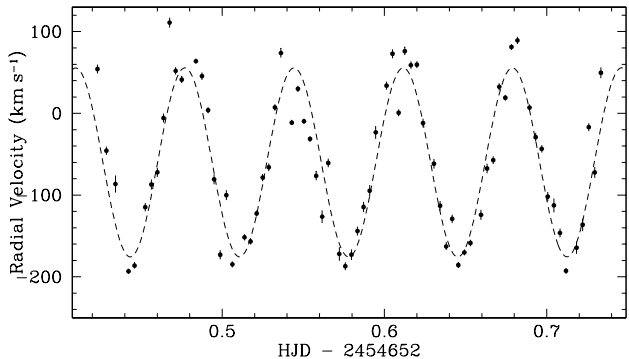}
\caption{\hel{ii}{5412} radial velocity curve obtained with WHT/ISIS on 2008 July 4. The velocities were measured by cross-correlating the individual profiles with a Gaussian template of $\mathrm{FWHM}=400$ \kms. A sine fit to the data results in a period of $0.06731\pm0.00038$\,d and an amplitude of $115\pm5~\kms$.}
\label{fig_longest_rvc}
\end{figure}

We obtained radial velocity curves of the \hel{ii}{5412} emission line on 2008 June 25, July 4, October 13, and October 14. By that time no \hel{ii}{4686} radial velocity variation was detected. This wasn't unexpected since the \hel{ii}{4686} line has to thin out (i.e. shed some nova ejecta emission) before it starts to present the same phenomenon as the weak, optically-thin transition of the \hel{ii}{5412} line. The \hel{ii}{4686} emission showed a clear modulation by November 2008. Therefore, we also measured radial velocities of this much brighter line on 2008 November 11, and 2009 May 25, July 21, and August 31. Before measuring the velocities, the spectra were first re-binned to constant velocity increments and continuum-normalised. Radial velocities were then measured by cross-correlation with a single Gaussian template. The FWHM of the template used for a given night was adjusted so that the cleanest radial velocity curve was obtained, but it always varied between 400 and 1200 \kms. The radial velocity curve of V458 Vul exhibits a quasi-sinusoidal
modulation. The longest observation (over 7\,h, 2008 July 4) covers over five cycles, and a sine fit to these data results in a period of $0.06731\pm0.00038$\,d and an amplitude of $115\pm5~\kms$ (Fig.~\ref{fig_longest_rvc}).

\begin{figure}
\includegraphics[width=\columnwidth]{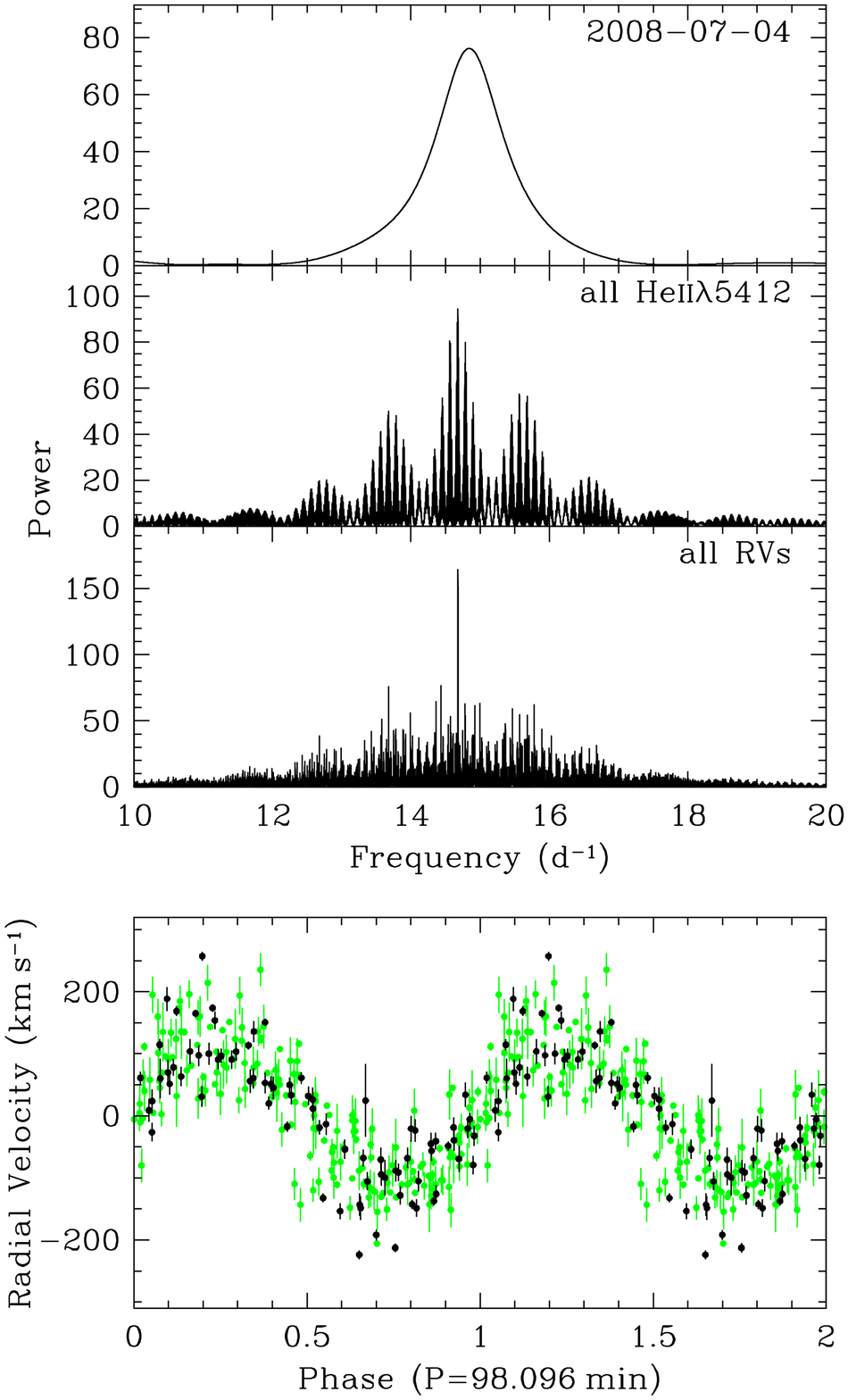}
\caption{{\it Top three panels}: \texttt{ORT} periodograms of the single longest observation (see Fig.~\ref{fig_longest_rvc}), of all the \hel{ii}{5412} radial velocities obtained between June and October 2008, and of all the \hel{ii}{5412} and \hel{ii}{4686} radial velocities obtained up to September 2009.  {\it Bottom panel}: Mean-subtracted and phase-folded \hel{ii}{5412} (green) and \hel{ii}{4686} (black) radial velocities. The time of zero phase (blue to red crossing) is $T_0 (\mathrm{HJD})=2454652.52694 \pm 8\times10^{-5}$.}
\label{fig_aov}
\end{figure}

In order to refine the orbital period of V458 Vul we subjected the radial velocity measurements to a period analysis using Schwarzenberg-Czerny's \citep{schwarzenberg-czerny96-1} variation of the analysis-of-variance method implemented as \texttt{ORT} in \texttt{MIDAS}, which fits periodic orthogonal polynomials to the phase-folded data. The periodogram calculated from the 2008 July 4 data (Fig.~\ref{fig_aov}, top panel) exhibits a strong peak at $14.86~\mathrm{d^{-1}}$, consistent with the result from the sine fit mentioned above. Next, we analysed the \hel{ii}{5412} radial velocities, which represent about 2/3 of all our radial velocity data and were obtained with relatively frequent sampling between June and October 2008. The resulting periodogram (Fig.~\ref{fig_aov}, middle panel) contains the strongest peak at $14.68~\mathrm{d^{-1}}$, and the observed alias pattern is consistent with the window function resulting from our temporal sampling. Finally, we analysed the combined \hel{ii}{5412} and \hel{ii}{4686} data, which extend the total baseline spanned by our observations to 430~d. The resulting periodogram is characterised by a narrow spike at $14.68~\mathrm{d^{-1}}$, consistent with period determinations of the smaller radial velocity subsets. No signal at the period claimed by \citeauthor{goranskijetal08-1} was found. A sine fit to the whole data set results in $P=0.06812255 \pm 0.00000017$~d or $98.09647\pm0.00025$~min. The \hel{ii}{5412} and \hel{ii}{4686} velocities folded on the orbital period are shown in Fig.~\ref{fig_aov} (bottom panel). Our results show that the period is coherent for 6341 cycles, suggesting it is a fixed clock in the system. We therefore identify this period with the orbital period of the binary progenitor of nova V458 Vul, which makes it the central binary system of a planetary nebula with the shortest orbital period \citep[see e.g.][for a list]{demarco09-1}.

\subsection{Trailed spectra diagrams}

\begin{figure*}
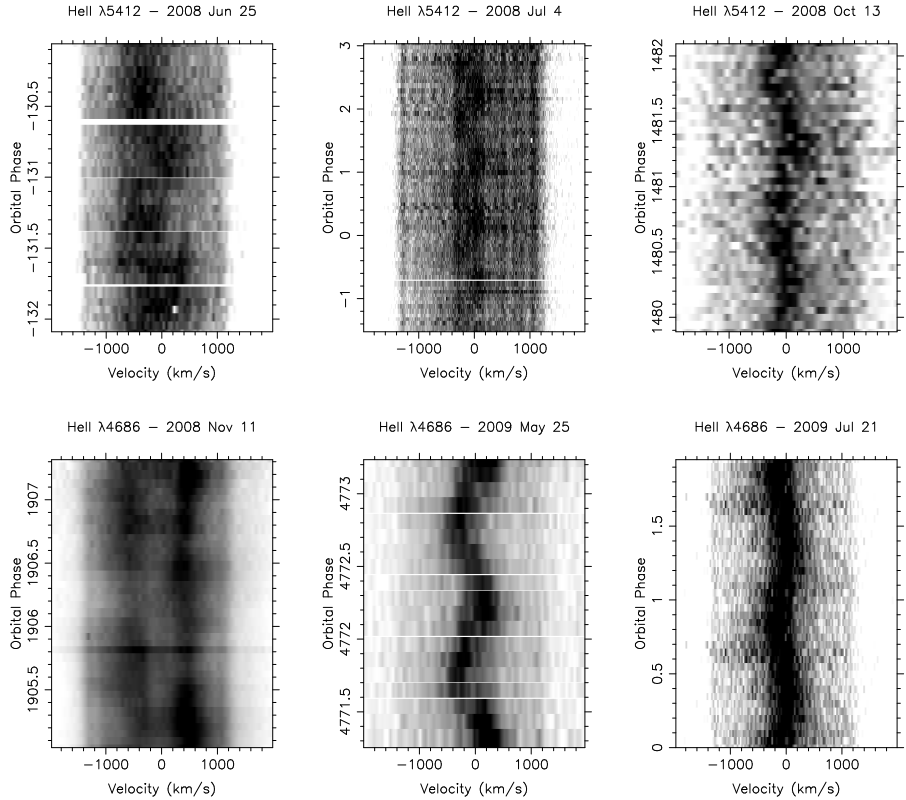

\includegraphics[width=5cm,angle=-90]{heii5412_trailed.ps}\\
\vspace{0.5cm}
\includegraphics[width=5cm,angle=-90]{heii4686_trailed.ps}
\caption{Evolution of the \he{ii} emission lines. {\em Top panel}: \hel{ii}{5412} trailed spectra diagrams. {\em Bottom panel}: \hel{ii}{4686} trailed spectra diagrams. Black represents emission. No phase binning has been applied, with the exception of the QUCAM data taken on 2009 July 21. In this last diagram a full cycle has been repeated for clarity.}
\label{fig_trailed}
\end{figure*}

The long-term evolution of the \hel{ii}{5412} and \hel{ii}{4686} emission lines is shown in Fig.~\ref{fig_trailed}. \hel{ii}{5412} started to reveal the orbital motion of V458 Vul much earlier than \hel{ii}{4686}. By May 2009, \hel{ii}{4686} displayed an apparent orbital signal in the form of a clear S-wave. Note that orbital phases were computed relative to the blue-to-red velocity crossing of this S-wave, which would correspond to the standard definition of the orbital phase if the S-wave originates on the donor star. The trailed spectra diagram of this line also shows high velocity wings extending up to $\sim \pm 1000$ \kms. This might indicate the presence of another emission component apart from the dominant S-wave.

\begin{figure}
\includegraphics[width=.63\columnwidth,angle=-90]{heii4686_trailed_May09.ps}
\includegraphics[width=.965\columnwidth]{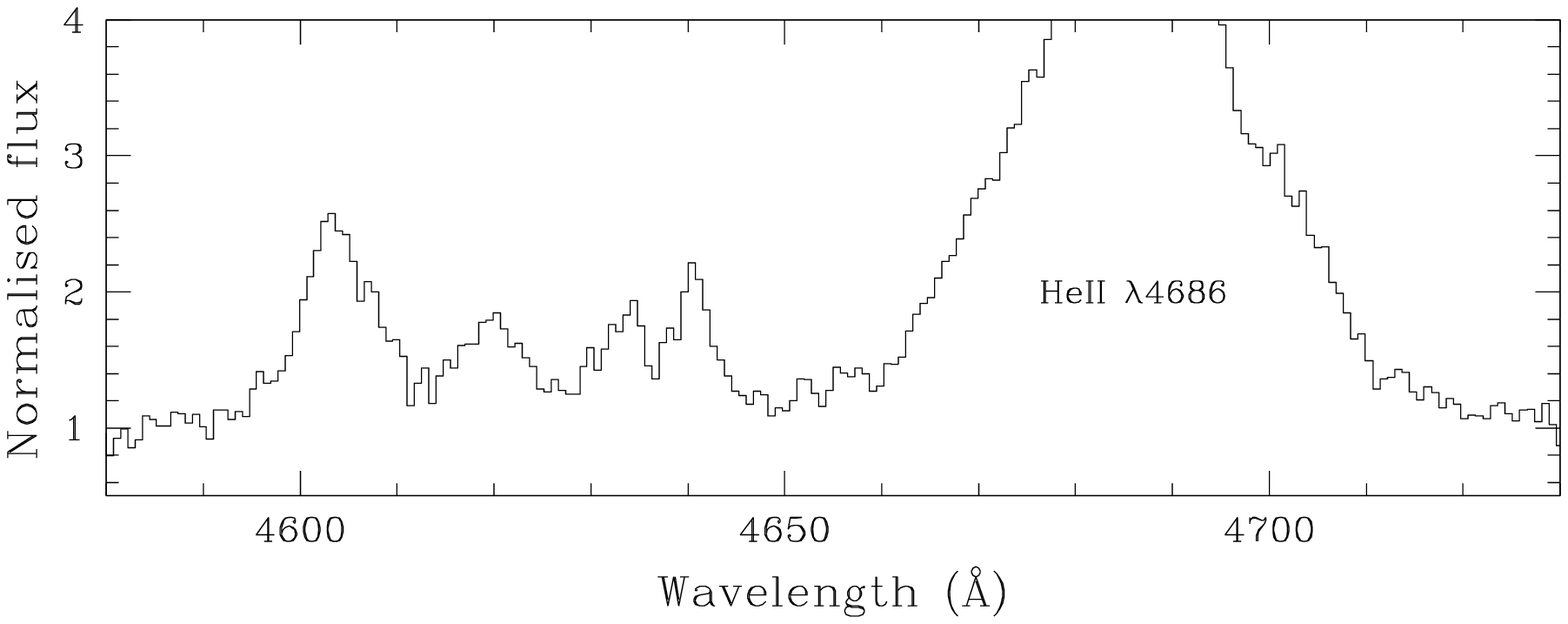}
\caption{{\em Top}: \hel{ii}{4686} trailed spectra diagram of the May 2009 run after averaging into 40 phase bins. Contrast has been adjusted to highlight the main S-wave ({\em left}) and the narrow components bluewards of \hel{ii}{4686} ({\em right}). Black represents emission and a full cycle has been repeated for clarity. {\em Bottom}: Doppler-corrected average spectrum of the May 2009 run.}
\label{fig_trailed2}
\end{figure}

A deeper look at the May 2009 spectra revealed narrow emission components bluewards of \hel{ii}{4686} (see Fig.~\ref{fig_trailed2}). The first two, counting from \hel{ii}{4686}, lie at rest wavelengths of $\sim 4640.6$ and $\sim 4634.2$ \AA~and have $\mathrm{FWHM} \sim 230$ \kms. These narrow lines, reminiscent of the radiation-driven Bowen fluorescence lines used to probe the motion of the irradiated donor star in X-ray binaries \cite[e.g.][]{steeghs+casares02-1}, are in phase with the \hel{ii}{4686} S-wave and their radial velocity amplitudes are comparable within what are necessarily substantial error bars. This lends further support to place these S-waves on the irradiated donor star. In V458 Vul, the white dwarf producing the nova explosion can provide the EUV radiation needed to trigger the process. In fact, two \ion{N}{iii} transitions take place at 4640.64 and 4634.13 \AA, very close to the observed lines. The other two emissions are likely the \ion{N}{v} doublet lines at 4603.74 and 4619.97 \AA. If all these narrow lines originate on the irradiated donor star, our adopted phase convention is the correct one.

\section{Discussion and conclusions}

In W08 we presented photoionisation modelling of the nebular spectra obtained before the nova explosion ionised the planetary nebula. This implied an ionising source with effective temperature $T_\mathrm{eff} \simeq 90000~\mathrm{K}$, luminosity $L_\mathrm{bol} \simeq 3000~\mathrm{L_\odot}$ and radius $R \simeq 0.23~\mathrm{R_\odot}$. In the same paper we showed that, based on the hydrogen-burning evolutionary tracks of \cite{vassiliadis+wood94-1}, this requires a core mass of $0.58~\mathrm{M_\odot}$ and an age since leaving the AGB consistent with our estimated nebular expansion age of 14000 years.

The question now is: which of the two stars is the progenitor of the planetary nebula? The short orbital period of V458 Vul and the age of its planetary nebula may seem at odds if one assumes the system is actually a cataclysmic variable (CV) which evolved from a much longer orbital period by losing angular momentum due to magnetic wind braking \citep{verbunt+zwaan81-1,rappaportetal83-1} and radiation of gravitational energy \citep{faulkner71-1,paczynski+sienkiewicz81-1}. The time it takes a CV to evolve down to an orbital period of 98.1 min is of the order of a Gyr \citep{rappaportetal83-1}. However, it is possible to get a short-period, {\it normal} CV within a common envelope \citep[e.g. the case of the young pre-CV SDSS J005245.11--005337.2 in][]{rebassa-mansergasetal08-1}, but producing a nova event in this scenario within 14000 years is very unlikely.   


This strengthens the possibility, as suggested by W08, of the donor star in V458 Vul being actually an evolved star, i.e. a post-AGB star. In such a case, the planetary nebula of V458 Vul may have been ejected by the donor star instead of the accreting white dwarf after a second common-envelope phase. As mentioned above, the post-AGB donor would therefore have a mass of $0.58~\mathrm{M_\odot}$.

Theoretical nova models \citep[e.g.][]{prialnik+kovetz95-1,yaronetal05-1} agree that a minimum white dwarf mass $M_1 \sim 1~\mathrm{M_\odot}$ is required to trigger the thermonuclear runaway in fast novae like V458 Vul. Observations, although scarce, point to a similar value \citep{ritter+kolb03-1}. Hence, the total mass of V458 Vul may well be $\gtrsim 1.6~\mathrm{M_\odot}$, above the critical Chandrasekhar mass, indicating that it may become a Type Ia supernova if the white dwarf manages to accumulate mass in the presence of nova eruptions.

Several other systems have been claimed as Type Ia supernova progenitors. The subdwarf-B+white dwarf binary KPD 1930+2752 is among the best candidates, but its total mass is very close to the critical mass \citep{maxtedetal00-1,ergmaetal01-1,geieretal07-1}. The first He nova, V445 Puppis, may contain a binary system composed of a massive white dwarf accreting from a helium star companion \citep{woudtetal09-1}. The 3.9-h central binary star of planetary nebula PNG135.9+55.9 (SBS 1150+599A) has also been put forward \citep{tovmassianetal10-1}. In this case, a post-AGB star and, presumably, a compact companion also amount to a mass just close to the Chandrasekhar limit.

An obvious objection to our scenario is the fact that the post-AGB donor star would have to fill its Roche lobe in order to sustain mass transfer while it is still contracting. A star filling its Roche lobe must obey an orbital period-mean density law, so we used the evolutionary tracks of \cite{bloecker95-1} in an attempt to find stellar parameters which fit both the 98-min orbit of V458 Vul and the results of our photoionisation model. We find that a star with an initial and final mass of $3~\mathrm{M_\odot}$ and $0.625~\mathrm{M_\odot}$, respectively, on a helium burning track at 14000 yr, provides almost perfectly the measured effective temperature, luminosity and radius of the ionising source. However, steady mass tranfer (i.e. contact with the Roche lobe) depends on the timescales over which orbital angular momentum is lost during the second common-envelope phase and contraction of the post-AGB donor star take place. Both processes have very short and similar timescales (of a few thousands years), making the situation very difficult to quantify. Only further spectroscopic search for spectral lines from both components of the binary system may shed more light onto its dynamics and nature.

In conclusion, we have solidly measured an orbital period of $98.09647 \pm 0.00025$ min for V458 Vul. A plausible scenario explaining V458 Vul is that of a double common-envelope binary system composed of a $M_1 \gtrsim 1~\mathrm{M_\odot}$ white dwarf (the accretor) and a $M_2 \sim 0.6~\mathrm{M_\odot}$, post-AGB star (the donor) which expelled the planetary nebula 14000 yr ago. The total mass of the system may therefore well exceed the Chandrasekhar mass which, in addition to its close orbit, makes V458 Vul a Type Ia supernova progenitor candidate.       
\section*{Acknowledgments}
We thank the referee, Nye Evans, for his comments. The use of Tom Marsh's \texttt{MOLLY} package is gratefully acknowledged. DS acknowledges a STFC Advanced Fellowship. The William Herschel Telescope and the Isaac Newton Telescope are operated on the island of La Palma by the Isaac Newton Group in the Spanish Observatorio del Roque de los Muchachos of the Instituto de Astrof\'\i sica de Canarias.

\bibliographystyle{mn2e}
\bibliography{/Users/prguez/Library/texmf/bib/mn-jour,/Users/prguez/Library/texmf/bib/aabib}

\label{lastpage}
\end{document}